\documentclass[a4paper,12pt]{article}
\usepackage{jcappub} 
\usepackage{cite}
\usepackage{graphicx}
\usepackage{enumerate}
\usepackage{hyperref}
\usepackage{cancel}
\usepackage{color}
\usepackage{graphicx}
\usepackage{amsmath}
\usepackage{amsfonts}
\usepackage{amssymb}
\usepackage{rotating}
\usepackage{booktabs}
\usepackage{xcolor}
\usepackage{soul}

\def\comment#1{}

\title{\boldmath 
Massive particle pair production and oscillation in Friedman Universe:
dark energy and matter interaction
}

\author{She-Sheng Xue}

\affiliation{ICRANet Piazzale della Repubblica, 10 -65122, Pescara, Italy, \\ Physics Department, University of Rome La Sapienza, \\ P.le Aldo Moro 5, I–00185 Rome, Italy
}

\emailAdd{xue@icra.it} 

\abstract{The classical Friedman equations of time-varying Hubble function $H$, dark-energy and matter densities couple to quantised field equations for massive modes $M\gg H$. Numerically solving these equations, we show the particle-antiparticle pairs production and oscillation in microscopic time scale ${\mathcal O}(M^{-1})$. A massive pair plasma state is formed in macroscopic time scale ${\mathcal O}(H^{-1})$. Its density and pressure introduce the interaction of matter and dark energy densities in the Friedman equations. Focusing on epochs after reheating, we show that the negative dark energy tracks down the radiation energy in the radiation epoch. Such tracking dynamics ends, and dark energy becomes positive in the matter epoch. The matter converts to dark energy, and their present values are comparable, explaining the cosmic coincidence. As a result, a class of effective interacting dark energy models is 
advocated to confront cosmological observations.}

\begin{document}
\maketitle
\flushbottom

%
\section{\bf Introduction}\label{int}

The Universe's evolution is gravitationally governed by matter and dark energy. The latter can be represented by the cosmological 
$\Lambda$ term in the Einstein equation. In addition to the mystery of 
its origin, people have not yet fully understood dark energy properties in Universe evolution.
In particular, how dark energy and matter interact with each other 
in Universe evolution and why their present values are coincidentally at 
the same order of magnitude. Such interacting dark energy can be 
simply represented by a time-varying cosmological $\tilde\Lambda(t)$ term 
in the Einstein equation or other modifications. 
Many theoretical ideas have been motivated for cosmology,
and advocated to examine the $H_0$ tension 
recently observed 
Refs.~\cite{DiValentino2017a,DiValentino2021b,Verde2019,DiValentino2021c,Freedman2017,Riess2019,Camarena2018,Salvatelli2013,Costa2014,Li2013,Yang2017,DiValentino2021,Yang2021,DiValentino2017,Zhao2017,Martinelli2019,Alestas2020,DiValentino2021a,Efstathiou2020,Yang2021a,Huang2016}. 
Here, we attempt to present a theoretical scenario to explain the dark energy and matter interaction by 
gravitational production and oscillation of particle-antiparticle
pairs via quantum back and forth reaction processes between dark energy $\tilde\Lambda(t)$ and massive pairs $M\gg H$. 

The gravitational particle production in Friedman Universe expansion is an 
important theoretical issue \cite{PhysRevLett.21.562,PhysRevD.3.346,PhysRev.183.1057,Birrell1984} that has been intensively studied for decades \cite{Mottola1985,Habib2000,Anderson2014,Anderson2014a,Landete2014}. Based on adiabaticity and non-back-reaction approximation for a slowly time-varying Hubble function $H(t)$, one adopted the semi-classical WKB approaches to calculating the particle production rate, which is exponentially suppressed $e^{-M/H}$ for massive particles $M\gg H$. However, the non-adiabatic back-reactions of particle creations on the Hubble function can be large and have to be taken into account.
The non-adiabatic back-reactions of massive particle productions have a quantum time scale ${\mathcal O}(1/M)$ that is much smaller than classical Universe evolution time scale ${\mathcal O}(1/H)$. 
To properly include the back-reaction of particle production on Universe evolution, one should separate fast components ${\mathcal O}(1/M)$ from slow components ${\mathcal O}(1/H)$ in the Friedman equation.
Many efforts 
\cite{Parker1973,Ford1987,Kolb1996,Greene1997,Kolb1998,Chung1998,Chung2001,Chung2000,Chung2003,Chung2005,Ema2016,Chung2019,Ema2018,Li2019,
Xue2019} have been made to study non-adiabatic back-reaction and 
understand massive particle productions without exponential suppression. 
It is important for reheating, possibly accounting for massive dark matter and total entropy of the present Universe \cite{Kofman1994,Kofman1997,Kuzmin1998,Kuzmin1999,Kolb1999,Bertone2005,Bassett2006,Kolb2007,Allahverdi2010,Frolov2010,Hall2010,Feng2010,Amin2014,Ema2015,Garny2016,Garny2018,Kolb2017,Hashiba2019,Hashiba2019a,Haro2019,Xue2020a}. 

In this article, we start with the Friedman equations for a flat Universe
\begin{eqnarray}
H^2=\frac{8\pi G}{3}\rho;\quad \dot H=-\frac{8\pi G}{2}(\rho + p),\label{friedman}
\end{eqnarray}
where energy density $\rho\equiv \rho_{_M}+\rho_{_\Lambda}$ and pressure 
$p\equiv p_{_M}+p_{_\Lambda}$. Equation of state $p_{_\Lambda}=-\rho_{_\Lambda}$ is for the cosmological term (dark energy), $p_{_M}=\omega_{_M}\rho_{_M}$ for the matter terms representing relativistic (radiation) and/or non-relativistic components. The second equation is a gernalised 
conservation law for time-varying cosmological term
$\rho_{_\Lambda}(t)\equiv \tilde\Lambda(t)/8\pi G$ \cite{Xue2015}, 
and it reduces to the usual equation 
$\dot \rho_{_M} + (1+\omega_{_M})H\rho_{_M}=0$ for time-constant $\rho_{_\Lambda}$.  
We adopt the 
approach \cite{Chung2019} to describe the decomposition of 
slow and fast components: scale factor $a=a_{\rm slow}+a_{\rm fast}$, Hubble function $H=H_{\rm slow}+H_{\rm fast}$, cosmological term and matter density $\rho_{_{\Lambda,M}}=\rho^{\rm slow}_{_{\Lambda,M}}+\rho^{\rm fast}_{_{\Lambda,M}}$ and pressure $p_{_{\Lambda,M}}=p^{\rm slow}_{_{\Lambda,M}}+p^{\rm fast}_{_{\Lambda,M}}$. The fast components vary much faster in time, but their amplitudes are much smaller than the slow components. According to the order of small ratio $\lambda$ of fast and slow components, the Friedman equations (\ref{friedman}) are decomposed into two sets. The slow components ${\mathcal O}(\lambda^0)$
obey the same equations as usual Friedman equations 
\begin{eqnarray}
H_{\rm slow}^2&=&\frac{8\pi G}{3}(\rho_{_M}^{\rm slow}+\rho_{_\Lambda}^{\rm slow});\nonumber\\
\quad \dot H_{\rm slow}&=&-\frac{8\pi G}{2}(\rho_{_M}^{\rm slow} +p_{_M}^{\rm slow}),
\label{sfriedman}
\end{eqnarray}
where $H_{\rm slow}=\dot a_{\rm slow}/a\approx 
\dot a_{\rm slow}/a_{\rm slow}$, 
time derivatives $\dot H_{\rm slow}$ and $\dot a_{\rm slow}$ 
relate to the macroscopic 
``slow'' time variation scale ${\mathcal O}(1/H)$.
The faster components ${\mathcal O}(\lambda^1)$ obey,
\begin{eqnarray}
H_{\rm fast}&=&\frac{8\pi G}{2\times 3H_{\rm slow}}(\rho_{_M}^{\rm fast}+\rho_{_\Lambda}^{\rm fast});\nonumber\\ 
\dot H_{\rm fast}&=&-\frac{8\pi G}{2}(\rho_{_M}^{\rm fast} +p_{_M}^{\rm fast}),
\label{ffriedman}
\end{eqnarray} 
where $H_{\rm fast}=\dot a_{\rm fast}/a\approx \dot a_{\rm fast}/a_{\rm slow}$, 
time derivatives $\dot H_{\rm fast}$ and 
$\dot a_{\rm fast}$ relate to the microscopic ``fast'' time variation scale ${\mathcal O}(1/M)$, and slow components are approximated as constants in ``fast'' time variation. For the cosmological term,
equation of state $p_{_\Lambda}=-\rho_{_\Lambda}$ becomes 
$p_{_\Lambda}^{\rm slow,fast}=-\rho_{_\Lambda}^{\rm slow,fast}$ respectively at order ${\mathcal O}(\lambda^0)$ 
and ${\mathcal O}(\lambda^1)$. In due course we shall clarify
the equation of state $p_{_M}=\omega_{_M}\rho_{_M}$ for the matter term.

We adopt the approach \cite{Parker1973} to describe the fast components of matter density $\rho_{_M}^{\rm fast}$ and pressure 
$p_{_M}^{\rm fast}$, that are attributed to the non-adiabatic production of particle and antiparticle pairs in fast time variation $H_{\rm fast}=\dot a_{\rm fast}/ a_{\rm slow}$. 
As a result, we find quantum pair production and oscillation and a macroscopic state of massive pair plasma. In 
radiation- and matter-dominated epochs after reheating,
we study how it affects the Friedman equation (\ref{sfriedman}) and 
introduces the interaction of dark-energy and matter densities. 
We show that the matter has converted to dark energy, and their present values are comparable, explaining the cosmic coincidence.    


\section{Quantum pair production and oscillation}\label{qppo}
A quantised massive scalar matter field inside the Hubble sphere volume 
$V\sim H^{-3}_{\rm slow}$ of Friedman Universe reads
\begin{eqnarray}
\Phi({\bf x},t)&=&\sum_n A_n Y_n({\bf x})\psi_n(t),
\label{qfield}
\end{eqnarray} 
which exponentially vanishes outside the horizon $H^{-1}_{\rm slow}$, and 
$\int_V Y_n({\bf x}) Y^\dagger_{n'}({\bf x})h^{1/2}d^3x=\delta_{nn'}$. 
The principal quantum number ``$n=0,1,2,\cdot\cdot\cdot$'' stands for  for quantum states of physical wave vectors $k_n$, $n=0$ and $k_0=0$ 
for the ground state \footnote{In Ref.~\cite{Parker1973}, the principal quantum number $n$ is the angular momentum number  ``$\ell=0,1,2,\cdot\cdot\cdot$'' and $Y_n({\bf x})=Y_{\ell,m}({\bf x})$ are the four-dimensional spherical harmonics for the closed Robertson-Walker metric. The ground state is $\ell =0$. 
}. 
The $A_n$ and $A_n^\dagger$ are time-independent annihilation and creation operators satisfying the commutation relation $[A_n^\dagger,A_n]=\delta_{n,n'}$. 
The time-separate equation for $\psi_n(t)$ is 
\begin{eqnarray}
\partial_t^2\psi_n(t) + \omega_n(t)^2\psi_n(t)=0,\quad 
\omega_n(t)^2=k^2_n+M^2,\label{timeeq}
\end{eqnarray} 
and Wronskian-type condition $\psi_n(t)\partial_t\psi^*_n(t) - \psi^*_n(t)\partial_t\psi_n(t)=i$. Expressing 
\begin{eqnarray}
\psi_n(t)\!&=&\!\frac{1}{(2V\omega_n)^{1/2}}\left(\alpha^*_n(t) e^{-i\int^t\omega_n dt}+\beta^*_n(t) e^{i\int^t\omega_n dt}\right)\label{alphabeta}
\end{eqnarray}
in terms of $\alpha_n(t)$ and $\beta_n(t)$, Equation (\ref{timeeq}) becomes
\begin{eqnarray}
\partial_t\alpha_n(t) &=& C_n e^{-2i\int^t\omega_n dt}\beta_n(t);\nonumber\\
 \partial_t\beta_n(t) &=& C_n e^{2i\int^t\omega_n dt}\alpha_n(t),
\label{eqalphabeta}
\end{eqnarray} 
and $|\alpha_n|^2-|\beta_n|^2=1$, where $C_n\equiv 3H\omega_n^{-2}[k_n^2/3 +M^2/2]$. 
In an adiabatic process for slowly time-varying $H=H_{\rm slow}$, the particle state
$\alpha_n(0)=1$ and $\beta_n(0)=0$ evolve to $|\alpha_n(t)|\gtrsim 1$ and
$|\beta_n(t)|\not=0$. Positive and negative frequency modes get mixed, 
leading to particle productions of probability $|\beta_n(t)|^2\propto e^{-M/H_{\rm slow}}$. 

We will focus on studying particle production in non-adiabatic processes 
for rapidly time-varying $H_{\rm fast}$, $\alpha_n$ and $\beta_n$ in the 
ground state $n=0$ of the lowest lying massive mode $M\gg H$. 
First, we recall that Parker and Fulling introduced the 
transformation \cite{Parker1973}, 
\begin{eqnarray}
A_0=\gamma^*B + \delta B^\dagger,\quad B=\delta A^\dagger_0-\gamma A_0,
\label{bogo}
\end{eqnarray}
$[B,B^\dagger]=1$, and two mixing constants 
obeying $|\gamma|^2-|\delta|^2=1$. For a given $A_n$ 
and its Fock space, the state $|{\mathcal N}_{\rm pair}\rangle $ 
is defined by the conditions 
$A_{n\not=0}|{\mathcal N}_{\rm pair}\rangle=0$ and  
\begin{eqnarray}
B^\dagger B |{\mathcal N}_{\rm pair}\rangle ={\mathcal N}_{\rm pair}|{\mathcal N}_{\rm pair}\rangle.
\label{pairB}
\end{eqnarray}
The $B^\dagger$ and $B$ are time-independent creation and annihilation 
operators of the pair of mixed positive frequency $A_0$ particle and 
negative frequency $A_0^\dagger$ antiparticle. 
The state $|{\mathcal N}_{\rm pair} \rangle $ 
contains ${\mathcal N}_{\rm pair}=1,2,3,\cdot\cdot\cdot$ pairs, 
and $|{\mathcal N}_{\rm pair}=0 \rangle $ is the ground state 
of non-adiabatic interacting system of fast varying $H_{\rm fast}$ 
and massive pair production and annihilation \footnote{Discussions can be applied for fermion fields. Analogously, we discussed the back and forth processes of massive fermion 
and antifermion pairs production and annihilation in spacetime 
${\mathcal S}\Leftrightarrow \bar F + F$ in Refs.~\cite{Xue2019,Xue2020a}}.
It is a coherent superposition of states of particle and anti-particle pairs. In this coherent condensate state $|{\mathcal N}_{\rm pair} \rangle$ and ${\mathcal N}_{\rm pair}\gg 1$, neglecting higher mode 
$n\not=0$ contributions, they obtained the negative quantum pressure 
and positive quantum density of 
coherent pair field, see Eqs.~(59) and (60) of Ref.~\cite{Parker1973},
\begin{eqnarray}
p^{\rm fast}_{_M}&=&-\frac{M(2{\mathcal N}_{\rm pair }+1)}{2\pi^2 V}\Big\{{\rm Re}[\gamma^*\delta(|\alpha|^2+|\beta|^2)]\nonumber\\
&+&(2|\delta|^2+1){\rm Re}(\alpha^*\beta e^{2iMt})\Big\},\label{fastp}\\
\rho^{\rm fast}_{_M}&=&\frac{M(2{\mathcal N}_{\rm pair }+1)}{\pi^2 V}\Big\{{\rm Re}[\gamma\delta^*\alpha\beta)]\nonumber\\
&+&(|\delta|^2+1/2)(|\beta|^2+1/2)\Big\},
\label{fastrho}
\end{eqnarray}
where $\omega_{n=0}=M$, $\alpha_{n=0}=\alpha$ and $\beta_{n=0}=\beta$.
Equations (\ref{fastp}) and density (\ref{fastrho}) were adopted for studying the avoidance of cosmic singularity in curved Universe. Note that $p^{\rm fast}_{_M}$ (\ref{fastp}) and $\rho^{\rm fast}_{_M}$ (\ref{fastrho})
represent the quantum pressure and density of massive coherent pair 
state (\ref{pairB}) in short quantum time sales ${\mathcal O}(1/M)$. They do not follow an usual 
equation of state of classical matter. 


Following their approach for the ground state $n=0$, we arrive at the same 
quantum pressure (\ref{fastp}) and density (\ref{fastrho}). 
We consider the state (\ref{pairB}) 
as a coherent condensate state of very massive $M\gg H_{\rm slow}$ and 
large number ${\mathcal N}_{\rm pair}\gg 1$ pairs, 
and $M(2{\mathcal N}_{\rm pair}+1)$ in (\ref{fastp}) and (\ref{fastrho}) 
can be larger than the Planck mass
$m_{\rm pl}$ so that higher mode 
$(n\not=0)$ contributions can be neglected. Moreover, 
we adopt (\ref{fastp}) and (\ref{fastrho}) as the fast 
components $\rho^{\rm fast}_{_M}$ and $p^{\rm fast}_{_M}$ in 
Eq.~(\ref{ffriedman}) to find their 
non-adiabatic back-reactions on fast components $H_{\rm fast}$ and 
$\rho^{\rm fast}_{_\Lambda}$. \comment{In the previous article, by using $M=10^{-2} m_{\rm pl}$, $H_{\rm slow}= 10^{-5} m_{\rm pl}$
and ${\mathcal N}_{\rm pair}\sim 10^{8}$, we show the massive pair productions and oscillations in inflation epoch, 
where the Hubble scale $H_{\rm slow}$ varies slowly.} 

\begin{figure*}[t]
\centering
\begin{center}
\includegraphics[height=5.5cm,width=7.8cm]{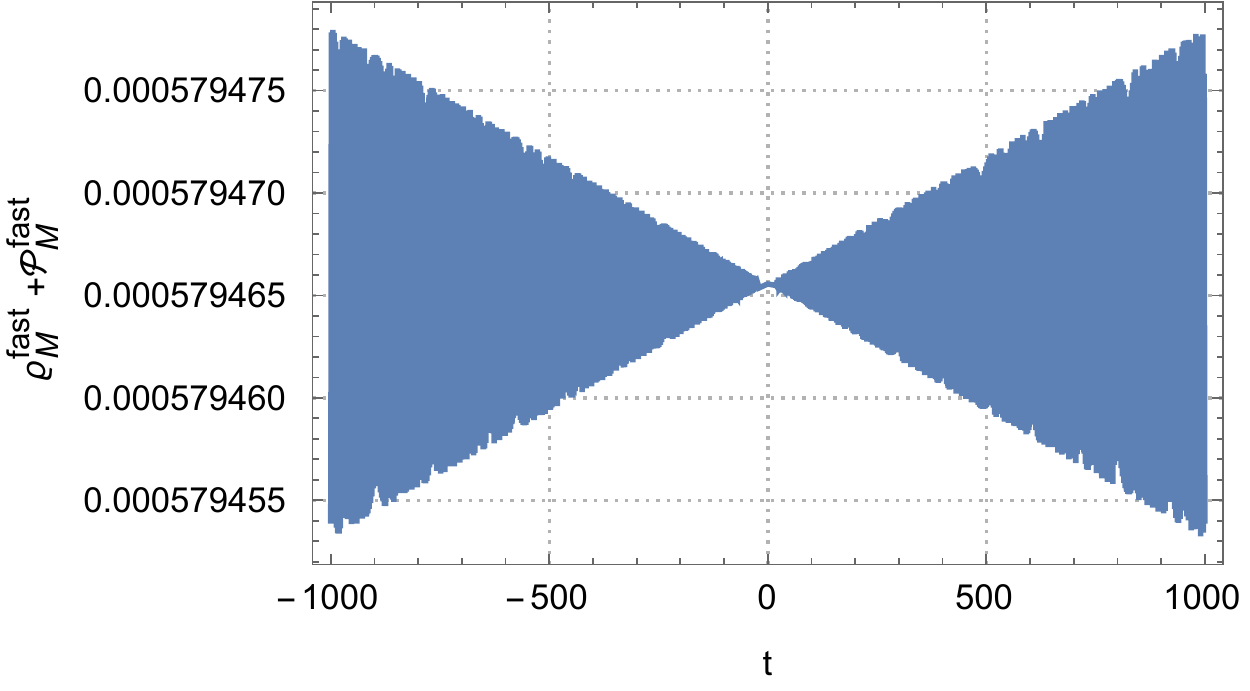}
\caption{We show the quantum pair density and pressure oscillations in microscopic time $t$ in unit of $M^{-1}$, using $H_{\rm slow}/M 
\approx 10^{-3}$, $M\simeq 10^{-10}m_{\rm pl}$, 
${\mathcal N}_{\rm pair}\simeq 10^{25}$ and $\delta= 1$. 
It is clear that for $H_{\rm slow}\ll m_{\rm pl}$ 
and $M\ll m_{\rm pl}$, a large amount of massive pairs 
${\mathcal N}_{\rm pair}\gg 1$ is created for significant oscillating quantum pressure (\ref{fastp+}) and density (\ref{fastrho+}). 
For details see Fig.~\ref{detailosci1+} 
in Supplemental Material. 
}\label{osci+}
\end{center}
\vspace{-2em}
\end{figure*}

Here, we study the epochs after reheating, when the Hubble scale and pair mass are very much smaller than the Planck mass, i.e., $H_{\rm slow}<M \ll m_{\rm pl}$ and ${\mathcal N}_{\rm pair}\gg 1$. Therefore, for a given $H_{\rm slow}$, 
we express in unit of the critical density $\rho_{\rm crit}=3m_{\rm pl}^2H^2_{\rm slow}$ the dimensionless quantum pressure (\ref{fastp}) and density (\ref{fastrho}) as 
\begin{eqnarray}
{\mathcal P}^{\rm fast}_{_M}&=&-\frac{\bar M H_{\rm slow}}{6\pi^2 m_{\rm pl}}\Big\{{\rm Re}[\gamma^*\delta(|\alpha|^2+|\beta|^2)]
+(2|\delta|^2+1){\rm Re}(\alpha^*\beta e^{2iMt})\Big\},\label{fastp+}\\
{\varrho}^{\rm fast}_{_M}&=&+\frac{\bar M H_{\rm slow}}{3\pi^2 m_{\rm pl}}\Big\{{\rm Re}[\gamma\delta^*\alpha\beta)]+(|\delta|^2+1/2)(|\beta|^2+1/2)\Big\},
\label{fastrho+}
\end{eqnarray}
where $\bar M\equiv (2{\mathcal N}_{\rm pair }+1)(M/m_{\rm pl})$ and the reduced Planck mass $m_{\rm pl}\equiv (8\pi G)^{-1/2}$.
The faster component equations (\ref{ffriedman}) become,
\begin{eqnarray}
h_{\rm fast}&=&\frac{1}{2}(\varrho_{_M}^{\rm fast}
+\varrho_{_\Lambda}^{\rm fast});\nonumber\\ 
\dot h_{\rm fast}&=&-\frac{3}{2}(\varrho_{_M}^{\rm fast} 
+{\mathcal P}_{_M}^{\rm fast}),
\label{ffriedman+}
\end{eqnarray}
where $h_{\rm fast}\equiv H_{\rm fast}/H_{\rm slow}$ and $\varrho_{_\Lambda}^{\rm fast}\equiv \rho_{_\Lambda}^{\rm fast}/\rho_{\rm crit}$.

Using negative ${\mathcal P}^{\rm fast}_{_M}$ (\ref{fastp+}) and positive definite $\varrho^{\rm fast}_{_M}$ (\ref{fastrho+}), we search for a solution of fast component equation (\ref{ffriedman+}) and quantum fluctuating mode equations (\ref{eqalphabeta}) in the period $[-t,t]$ of the microscopic time $t\sim H^{-1}_{\rm fast}$, which is around the macroscopic time $t_{\rm slow}\sim H^{-1}_{\rm slow}$, when the slow components $a_{\rm slow}$, $H_{\rm slow}$, $\rho^{\rm slow}_{_{M,\Lambda}}$ and $p^{\rm slow}_{_{M,\Lambda}}$ are valued, following 
the Friedman equations (\ref{sfriedman}).  The integrals 
$\int^t\omega_n dt$ are over the microscopic time $t$  characterised by 
the Compton time scale $1/M$. Its lower limit is $t=0$ by setting 
$t_{\rm slow}=0$ as a reference time, when $a_{\rm fast}(0)=0$,
\begin{eqnarray}
H_{\rm fast}(0)=\dot a_{\rm fast}/a_{\rm slow}=0;\quad \alpha(0)=1,\quad \beta(0)=0.
\label{initial}
\end{eqnarray}
The real value $\gamma^*\delta$ condition in Eqs.~(\ref{fastp+}) 
and (\ref{fastrho+}) leads to the time symmetry:
$a^{\rm fast}(t)=a^{\rm fast}(-t)$, $\alpha(t)=\alpha^*(-t)$ 
and $\beta(t)=\beta^*(-t)$ \cite{Parker1973}. When $t\leftrightarrow -t$, positive and negative frequency modes interchange. 
\comment{In Ref.~\cite{Parker1973}, $a_{\rm slow}=0$, $H_{\rm slow}=0$ (i.e.,
$H=H_{\rm fast}$, $a=a_{\rm fast}$) and a small spherical volume $V\sim (H_{\rm fast})^3$ at the cosmic origin were adopted for studying the avoidance of cosmic singularity for the $\rho_{_\Lambda}=0$ and curved Universe.} 
Here we use $a_{\rm slow}\not=0$, $H_{\rm slow}\not=0$ and co-moving radius $(Ha)^{-1}\approx (H_{\rm slow}a_{\rm slow})^{-1}$ of Hubble volume $V\sim H_{\rm slow}^{-3}$.

In microscopic time $t$ of unit $M^{-1}$, we numerically solve coupled Eqs.~(\ref{eqalphabeta}) and (\ref{fastp+}-\ref{ffriedman+}) with the initial condition (\ref{initial}). 
Figure \ref{osci+} shows results for $C_0 = (3/2)h_{\rm fast}(H_{\rm slow}/M)$ and verified condition $|\alpha|^2-|\beta|^2=1$. 
In the quantum period of microscopic time $t$, the negative quantum pressure 
${\mathcal P}^{\rm fast}_{_M}< 0$ and back-reaction effects lead to the {\it quantum pair oscillation} characterised by the frequency $\omega=M$ of massive quantised pair fields. The positive quantum pair density $\varrho^{\rm fast}_{_M}> 0$ indicates particle creations without $e^{-M/H}$ suppression. 
It is consistent with increasing Bogoliubov 
coefficient $|\beta(t)|^2$ that mixes positive and negative energy modes. 
Observe that $\varrho^{\rm fast}_{_M}\gg |{\mathcal P}^{\rm fast}_{_M}|$ and the sum $\varrho^{\rm fast}_{_M}+ {\mathcal P}^{\rm fast}_{_M} >0$ is positive definite, leading to the decreasing $h_{\rm fast}(t)$ (\ref{ffriedman}). As a consequence,   
for time $t>0$, the fast components $h_{\rm fast}$ 
and $\varrho^{\rm fast}_{_\Lambda}$ decrease in time, in order for pair production. Whereas for time $t<0$, $h_{\rm fast}$ and $\varrho^{\rm fast}_{_\Lambda}$ increases, due to pair annihilation. The small $a_{\rm fast}(t)$ varies around $a_{\rm slow}$ at $t_{\rm slow}\equiv 0$. 

The quantum pair oscillation phenomenon is dynamically analogous 
to the plasma oscillation of electron-positron pair production in 
an external electric filed $E$ \cite{Kluger1991} and pair production rate 
is not exponentially suppressed by $e^{-\pi M^2/E}$ \cite{Ruffini2010}. 
The coherent plasma state of electron-positron pairs is analogous to the 
coherent pair state $|{\mathcal N}_{\rm pair}\rangle $ (\ref{pairB}).

\section{Massive pair plasma state}\label{mpp}

As shown in Fig.~\ref{osci+}, massive pair quantum pressure ${\mathcal P}^{\rm fast}_{_M}$ (\ref{fastp+}) and density $\varrho^{\rm fast}_{_M}$ (\ref{fastp+})  can be significantly large and rapidly oscillate with the fast components $h_{\rm fast}$ and $\varrho^{\rm fast}_{_\Lambda}$ (\ref{ffriedman+}) in microscopic time and space. Their oscillating amplitudes are not dampen in time, and it is therefore expected to form 
{\it a massive pair plasma state} in a long macroscopic time. 
However, to study their effective impacts on the classical Friedman equations (\ref{sfriedman}) evolving in macroscopic time and space, we have to discuss two problems coming from scale difference 
$M\gg H_{\rm slow}$. First, it is impossible to even numerically integrate slow and fast component coupled equations (\ref{sfriedman},\ref{ffriedman}) due to their vastly different time scales. On this aspect, we consider their 
non-vanishing averages $\langle\cdot\cdot\cdot\rangle$ over the microscopic period in time. Figure \ref{osci+} shows $\langle\varrho^{\rm fast}_{_M}+ {\mathcal P}^{\rm fast}_{_M}\rangle$ and other averages of fast oscillating components do not vanish. Second the spatial dependence of pair quantum pressure ${\mathcal P}_{\rm fast}$ (\ref{fastp}) and density $\varrho_{\rm fast}$ (\ref{fastp}) are unknown, since they are obtained 
by using the vacuum expectation value of field $\Phi({\bf x},t)$ energy-momentum tensor over entire space. For the case $M\gg H_{\rm slow}$, 
the Compton length $M^{-1}$ of ground state $n=0$ is much smaller than the Hubble horizon $H^{-1}_{\rm slow}$. Therefore, the massive coherent pair state (\ref{pairB}-\ref{fastrho}) and quantum plasma oscillation of Fig.~\ref{osci+} well localise inside the Hubble sphere. We speculate that their location should be nearby the horizon because of isotropic homogeneity extending up to the horizon.  

Based on these considerations and non-vanishing averages 
of fast oscillating components (Fig.~\ref{osci+}) over macroscopic 
time, we assume the formation of massive pair plasma state in macroscopic time scale. We describe such macroscopic state as a perfect fluid state of effective number $n^H_{_M}$ and 
energy $\rho^H_{_M}$ densities as,
\begin{eqnarray}
\rho^H_{_M} \equiv  2\chi  m^2 H^2_{\rm slow},\quad n^H_{_M} \equiv   \chi  m H^2_{\rm slow};\quad m^2 \equiv 
\sum_fg_d^fM^2_f,
\label{apdenm}
\end{eqnarray}
and pressure $p^H_{_M}=\omega^H_{_M}
\rho^H_{_M}$. The $\omega^H_{_M}\approx 0$ for $m\gg H_{\rm slow}$
and its upper limit is $1/3$. The introduced mass parameter $m$ represents possible particle masses $M_f$, degeneracies $g_d^f$ 
and the mixing coefficient $\delta$ (\ref{bogo}). 
The degeneracies $g_d^f$ plays the same role of pair number ${\mathcal N}_{\rm pair}$ in 
Eq.~(\ref{fastrho}), namely $\sum_f g_d^f\approx (2{\mathcal N}_{\rm pair}+1)$. We explain 
the reasons why the densities (\ref{apdenm}) 
are proportional to $\chi m H^2_{\rm slow}$, rather than $H^3_{\rm slow}$ 
from the entire Hubble volume $V$. The ``surface area'' factor $H^2_{\rm slow}$ 
is attributed to the 
spherical symmetry of Hubble volume. The ``radial size'' factor $\chi m$ comes from 
the layer width $\lambda_m $ introduced as an effective parameter to
describe the properties (i) for $m\gg H_{\rm slow}$ the massive pair plasma
is localised as a spherical layer and (ii) its radial width $\lambda_m < H^{-1}_{\rm slow}$
depends on the massive pair plasma oscillation dynamics \footnote{It may also include self-gravitating dynamics, due to pair plasma are very massive.}, 
rather than the $H_{\rm slow}$ dynamics govern by the Friedman equations (\ref{sfriedman}). 
The width parameter $\chi$ expresses the layer width $\lambda_m = (\chi m)^{-1} \gg 1/m$ 
in terms of the effective Compton length $1/m$,
\begin{eqnarray}
\lambda_m=(\chi m)^{-1} < H^{-1}_{\rm slow},\quad  1\gg \chi > (H_{\rm slow}/m).
\label{chi}
\end{eqnarray} 
Because parameters $m$ and $\chi m$ represent time-averaged values over 
fast time oscillations of massive pair plasma state, we consider $m$ and 
$\chi m$ as approximate constants in slowly varying macroscopic 
time. However, the $m$ and $\chi m$ values, namely the $M_f$ and $g_d^f$ values (\ref{apdenm}) cannot be unique in entire Universe evolution, and should depend on Universe evolution epochs. One of the reasons is the fast-component equations for massive pair productions and oscillations depend on the $H_{\rm slow}$ value, see Sec.~\ref{qppo}. We will duly come back to this point how characteristic value $\chi m$ relates to 
the Hubble function $H_{\rm slow}$ in a given evolution epoch. 

We have to point out that (i) the pressure $p^H_{_M}$ and density $\rho^H_{_M}$ (\ref{apdenm}) are effective descriptions of the massive pair plasma state in macroscopic scales, that may result from the coherence condensation state (\ref{pairB},\ref{fastp},\ref{fastrho}) and oscillating dynamics (Fig.~\ref{osci+}) in microscopic scales; (ii) they play the role of ``slow'' components contributing to the Friedman equations (\ref{friedman}) or (\ref{sfriedman}). It means that in the Friedman equations, there are two sets of the matter: (i) the normal matter state of pressure and density $p_{_M}=\omega_{_M}\rho_{_M}$ and (ii) the massive pair plasma state of pressure and density 
$p^H_{_M}=\omega^H_{_M}\rho^H_{_M}$. These two sets interact with each other, shown below. We shall study the massive pair plasma state effects on each epoch of Universe evolution. Here we start to study its effects on the epoch after reheating. Henceforth sub- and super-scripts ``slow'' are dropped.    

\section{Cosmic rate equation}\label{mppa}

Up to macroscopic time $H^{-1}$, we estimate the total number of particles produced inside the Hubble sphere $N\approx n^H_{_M}H^{-3}/2$ and mean pair production rate w.r.t.~macroscopic time
\begin{eqnarray}
\Gamma_M \approx \frac{dN}{2\pi dt}\approx \frac{\chi m}{4\pi} \epsilon,\quad \epsilon\equiv -\frac{\dot H}{H^2}=\frac{3}{2}\frac{(1+\omega_{_M})\rho_{_M}}{\rho_{_\Lambda}+\rho_{_M}}.
\label{prate}
\end{eqnarray}
It is not a theoretical derivation, but modelling parameterized 
by $\chi m$ and Universe evolution rate $\epsilon$. The asymptotic values $\epsilon \approx  2$ and $\epsilon \approx  3/2$ are respectively for radiation and matter epoch.
Here we neglect 
the back-reactions of slow time-varying components $H$, $\rho_{_{\Lambda,M}}$ and $p_{_{\Lambda,M}}$ on fast components $H_{\rm fast}$, $\rho^{\rm fast}_{_M}$ and $p^{\rm fast}_{_M}$.

We turn to study how the massive pair plasma density interacts with the matter density $\rho_{_M}$ that governs the Universe evolution, 
\begin{eqnarray}
\rho^H_{_M}\Leftrightarrow \rho_{_M}.
\label{bfrho}
\end{eqnarray}
Recall the rate equation for the back and forth process $e^+e^-\Leftrightarrow\gamma\gamma$ \cite{Kolb1990,Lee,Ruffini1999,Ruffini2000}:
\begin{eqnarray}
\frac{dn_{e^+e^-}(t)}{dt} +3 H n_{e^+e^-}(t) = \langle\sigma v \rangle \Big(n^2_{e^+e^-}\big|_{\rm eq}-n^2_{e^+e^-}\Big),
\label{rateee}
\end{eqnarray}
where $n_{e^+e^-}(t)$ is the density governed by macroscopic evolution and $n_{e^+e^-}\big|_{\rm eq}$ is the density in an equilibrium with photons. The RHS represents the averaged interacting rate $dN/dt\approx \langle\sigma v \rangle n_{e^+e^-}$ for microscopic detail balance between $n_{e^+e^-}$ and $n_{e^+e^-}\big|_{\rm eq}$. They are coupled for $n_{e^+e^-}\big|_{\rm eq}\approx n_{e^+e^-}$ and decoupled $n_{e^+e^-}\big|_{\rm eq}\ll n_{e^+e^-}$. This motivates us to propose an effective cosmic rate equation, 
\begin{eqnarray}
\dot\rho_{_M}+ 3(1+\omega_{_M}) H\rho_{_M} &=& \Gamma_M(\rho_{_M}^H - \rho_{_M}) 
\label{rateeqd}
\end{eqnarray}
for the the back and forth  $\rho_{_M}$ and $\rho^H_{_M}$ interaction (\ref{bfrho}) in the Universe evolution. It actually represents a general conservation law of all matter including massive pair plasma density 
$\rho_{_M}^H$ (\ref{apdenm}) with the production rate (\ref{prate}). 
The term $3(1+\omega_{_M}) H\rho_{_M}$ of the time scale $(3H)^{-1}$ represents the space-time expanding effect on the density $\rho_{_M}$. While $\Gamma_M \rho_{_M}^H$ is the source term and 
$\Gamma_M\rho_{_M}$ is the depletion term. The time-varying horizon $H$ and massive pair plasma state are coupled via the back and forth processes (\ref{bfrho}). The ratio $\Gamma_M/H> 1$ indicates the coupled case, and  $\Gamma_M/H < 1$ indicate the decoupled case. 

We see how the massive pair plasma density (\ref{apdenm}) and cosmic rate equation (\ref{rateeqd}) affect on the Friedman equations (\ref{sfriedman}). The cosmic rate equation (\ref{rateeqd}) combines with  
Eqs.~(\ref{sfriedman}), yielding 
\begin{eqnarray}
\dot\rho_{_\Lambda}&=& -  \Gamma_M \left(\rho^H_{_M} - \rho_{_M}\right).
\label{rhol}
\end{eqnarray}
Equations (\ref{rateeqd}) and (\ref{rhol}) is reminiscent of a generally modeling interacting dark energy and matter $\delta Q=\Gamma_M(\rho_{_M}^H - \rho_{_M})$, based on the total mass-energy conservation, 
see review \cite{Wang2016,DiValentino2020} and \cite{Guo2017,Feng2019,Guo2018}. It shows that the cosmological constant (dark energy) 
$\rho_{_\Lambda}$ and matter energy $\rho_{_M}$ interact via the massive pair plasma
$\rho^H_{_M}$ produced by massive particle production and oscillation in the Friedman space. Two cases: (i) dark energy converts to matter energy when $\rho_{_M}^H > \rho_{_M}$ and  (ii) matter energy converts to dark energy when $\rho_{_M}^H < \rho_{_M}$. 

Equations (\ref{sfriedman}) and (\ref{rateeqd}) are a set of first-order ordinary  differential equations, numerical solutions for $\rho_{_M}$ 
and $\rho_{_\Lambda}$ can be studied, provided that initial or transition conditions from one epoch to another are known. In this article, 
we approximately find asymptotic solutions of specific epochs to gain a qualitative insight into how dark energy and matter interact in Universe evolution. 

\section{Radiation and matter dominate epochs}

Suppose that all radiation $\rho_{_R}$ and matter $\rho_{_M}$ densities were created in the reheating epoch, and they were much larger than the dark energy density 
$\rho_{_\Lambda}$ \footnote{The reheating epoch is discussed in a separated article 
Ref.~\cite{Xue2020a}}. The standard cosmology started with the radiation dominated epoch and proceeded to the matter-dominated epoch. We adopt an analytical way to reveal approximate $\rho_{_\Lambda}-\rho_{_R}$ and $\rho_{_\Lambda}-\rho_{_M}$ relations.

In radiation dominate epoch, we replace $\rho_{_M} \rightarrow \rho_{_R}$ and $\omega_{_M}\rightarrow \omega_{_R}\approx 1/3$ in Eqs.~(\ref{rateeqd}) and (\ref{rhol}).  Neglecting dark-energy and non-relativistic matter densities 
$H^2\approx \rho_{_R}/(3m^2_{\rm pl})$, 
and $\rho_{_M}^H\approx (2\chi/3)\bar m_{_R}^2\rho_{_R}$, we recast Eqs.~(\ref{rateeqd}) and (\ref{rhol}) as  
\comment{
\begin{eqnarray}
\frac{d\rho_{_R}}{dx} + 4 \rho_{_R} &=&  + \frac{3^{1/2}\chi \bar m_{_R}m^2_{\rm pl}}{6\pi}[\chi\bar m_{_R}^2-3]\rho_{_R}^{1/2}
\label{rhomr}\\
\frac{d\rho_{_\Lambda}}{dx} &=& - \frac{3^{1/2}\chi \bar m_{_R}m^2_{\rm pl}}{6\pi}[\chi\bar m_{_R}^2-3]\rho_{_R}^{1/2}
\label{rholr}
\end{eqnarray}
}
\begin{eqnarray}
\frac{d\rho_{_R}}{dx} + 4 \rho_{_R} &=&  + \langle \Gamma_M/H\rangle_{_R}
[\chi\bar m_{_R}^2-1]\rho_{_R}
\label{rhomr}\\
\frac{d\rho_{_\Lambda}}{dx} &=& - \langle \Gamma_M/H\rangle_{_R} 
[\chi\bar m_{_R}^2-1]\rho_{_R}
\label{rholr}
\end{eqnarray}
by using the new variable $x=\ln a$ and $dx=Hdt$. Because the  radiation epoch is very long and the $H$ varies a lot, the mass $m$ 
(\ref{apdenm}) and width parameter $\chi m\propto H$ (\ref{chi}) vary as well, which we cannot go to details. Thus we introduce the average mass 
parameter $m_{_R}=\langle m\rangle_{_R}$ and average rate $\langle \Gamma_M/H\rangle_{_R}$ over the entire radiation epoch, assuming they vary much slowly than $\rho_{_R}$ and $\rho_{_\Lambda}$. The dimensionless average mass parameter $\bar m_{_R}\equiv (2/3)m_{_R}/m_{\rm pl}$ and $\chi \bar m_{_R}^2 < 1$. 

The asymptotic solutions are
\comment{
\begin{eqnarray}
\rho_{_R} &\approx & \Big[ \tilde C_{_R}\left(\frac{a_{_R}}{a}\right)^2-\frac{3^{1/2}\chi\bar m_{_R} m^2_{\rm pl}}{4\pi} \Big]^2 \approx  \rho^{\rm RH}_{_R}\left(\frac{a_{_R}}{a}\right)^{4-\delta^G_{_M}}\\
\rho_{_R} &\approx&  \rho^{\rm RH}_{_R}\left(\frac{a_{_R}}{a}\right)^4;\quad 
\rho_{_\Lambda} \approx - \frac{3^{3/2}\chi \bar m_{_R}m^2_{\rm pl}}{4\pi} \rho_{_R}^{1/2} +\tilde C_{_\Lambda}. 
\label{rholrs}
\end{eqnarray}
}
\begin{eqnarray}
\rho_{_R} &=&  \rho^{\rm RH}_{_R}\left(\frac{a_{_R}}{a}\right)^{4-\gamma_{_R}}, \quad \gamma_{_R}\equiv 
\langle \Gamma_M/H\rangle_{_R}(\chi\bar m^2_{_R}-1)\label{rhors}\\
\rho_{_\Lambda} &=&  \frac{\gamma_{_R}}{4-\gamma_{_R}} \rho_{_R} +\tilde {\mathcal C}_{_\Lambda}, \quad \rho_{_\Lambda} = \frac{\gamma_{_R}}{4-\gamma_{_R}} \rho_{_R}. 
\label{rholrs}
\end{eqnarray}
The dark-energy and matter coupling parameter $\gamma_{_R} < 0$ ($|\gamma_{_R}|< 1$) 
represents the $\rho_{_\Lambda}-\rho_{_R}$ interaction and $\rho_{_R}$ 
conversion to $\rho_{_\Lambda}$. 
The initial values $\rho^{\rm RH}_{_R}$ and $\tilde C_{_\Lambda}$ are given 
at the reheating end $a=a_{_R}$. In this article, to study dark energy and radiation interaction, we select the initial condition 
$\tilde C_{_\Lambda} =0$, consistently with 
$\rho^{\rm RH}_{_\Lambda}\propto\rho^{\rm RH}_{_R}$ and $\rho^{\rm RH}_{_\Lambda}\ll 
\rho^{\rm RH}_{_R}$ at reheating end. 
The reasons are that the dark energy $\rho_{_\Lambda}$ converts to massive pair plasma energy $\rho^H_{_M}$ (\ref{apdenm}), and massive pairs decay to relativistic particles, producing  radiation energy $\rho_{_R}$ \cite{Xue2020a}. With such an initial condition $\tilde C_{_\Lambda} =0$, the dark-energy in radiation epoch is negative $\rho_{_\Lambda}<0$ because of 
$\gamma_{_R} < 0$. The detailed discussions about negative dark energy can be found in Refs.~\cite{Calderon2021,Boisseau2015,Vazquez2012,Ye2020,Akarsu2019,Dutta2018,Grande2006}.  
It requires more studies of the transition from reheating to radiation epochs to determine 
$\tilde C_{_\Lambda}$. 

As a result, the asymptotic solution (\ref{rholrs}) shows that 
$\rho_{_\Lambda}$ linearly tracks down (follows) $\rho_{_R}$. Here we adopt the terminology ``track down'' used in the discussions of 
Ref.~\cite{Zlatev1999}. Such $\rho_{_\Lambda}-\rho_{_R}$ tracking dynamics continues in the entire radiation epoch. We will show that 
the tracking dynamics ends and $\rho_{_\Lambda}$ becomes positive 
during a continuous transition period from radiation epoch to matter epoch. 
We use an analytical approach to asymptotic solutions in different epochs. Therefore we cannot precisely determine the transition period.
Therefore, we introduce the scale factor $a_{\rm tr}$ to characterize the transition time scale, and discuss two extremal cases:
\begin{enumerate}[(i)] 
\item  $a_{\rm tr}\sim a_{\rm eq}$ transition occurs at the radiation-matter equilibrium moment; 
\item  $a_{\rm tr} > a_{\rm eq}$ transition occurs at sometime around/after the last scattering surface.  
\end{enumerate}
More details of the transition behave and period need numerical studies of massive pair plasma (\ref{apdenm},\ref{prate}), Friedman equation (\ref{sfriedman}) and cosmic rate equation (\ref{rateeqd}). 


In the matter dominate epoch, we identify $\rho_{_M} \rightarrow \rho_{_M}$ and $\omega_{_M}\rightarrow \omega_{_M}\approx 0$ in Eqs.~(\ref{rateeqd}) and (\ref{rhol}).
Analogously to the approach in radiation epoch, neglecting dark-energy and radiation-energy density, $H^2\approx \rho_{_M}/(3m^2_{\rm pl})$, 
and $\rho_{_M}^H\approx \chi\bar m_{_M}^2\rho_{_M}$, 
we recast Eqs.~(\ref{rateeqd}) and (\ref{rhol}) as  
\begin{eqnarray}
\frac{d\rho_{_M}}{dx} + 3 \rho_{_M} &=&  + \langle \Gamma_M/H\rangle_{_M} 
(\chi\bar m_{_M}^2-1)\rho_{_M},
\label{rhomm}\\
\frac{d\rho_{_\Lambda}}{dx} &=& - \langle \Gamma_M/H\rangle_{_M}
(\chi\bar m_{_M}^2-1)\rho_{_M},
\label{rholm}
\end{eqnarray} 
where $\bar m_{_M}\equiv (2/3)m_{_M}/m_{\rm pl}$ and $\chi \bar m_{_M}^2< 1$. Here we introduce the average mass parameter $m_{_M}$ and rate 
$\langle \Gamma_M/H\rangle_{_M}$ over the matter epoch, assuming they vary much slowly than $\rho_{_M}$. 
The asymptotic solutions are
\begin{eqnarray}
\rho_{_M} &=& \rho^{\rm eq}_{_M}\left(\frac{a_{\rm eq}}{a}\right)^{3-\gamma_{_M}},\quad \gamma_{_M}\equiv \langle \Gamma_M/H\rangle_{_M}
(\chi\bar m^2_M-1),\label{rhoms}\\
\rho_{_\Lambda} &= &\frac{\gamma_{_M}}{3-\gamma_{_M}} \rho_{_M} + \tilde {\mathcal C}^{\rm eq}_{_\Lambda},\quad \rho_{_\Lambda}\rightarrow \rho^0_{_\Lambda}\approx \tilde {\mathcal C}^{\rm eq}_{_\Lambda}. 
\label{rholms}
\end{eqnarray}
The coupling parameter $\gamma_{_M} < 0$ ($|\gamma_{_M}|< 1$) 
represents the $\rho_{_\Lambda}-\rho_{_M}$ interaction and $\rho_{_M}$ 
conversion into $\rho_{_\Lambda}$. Here we adopt the case (i) $a_{\rm tr}\sim a_{\rm eq}$ for discussions. Namely, the $\rho_{_\Lambda}-\rho_{_R}$ tracking continues until the Universe reaches the radiation-matter equilibrium $\rho^{\rm eq}_{_M}=\rho^{\rm eq}_{_R}$ 
at $(a_{\rm eq}/a_{_R}) = (T_{\rm RH}/T_{\rm eq})\sim 10^{15}{\rm GeV}/ 10\,{\rm eV}\sim 10^{23}$, where $T_{\rm RH} (T_{\rm eq})$ is the reheating (equilibrium) temperature.  The initial value $\rho^{\rm eq}_{_M}$
is given at the radiation-matter equilibrium $\rho^{\rm eq}_{_M}=\rho^{\rm eq}_{_R}$ 
at $a=a_{\rm eq}$, where the solutions (\ref{rholrs})
and (\ref{rholms}) should match, yielding
\begin{eqnarray}
\tilde {\mathcal C}^{\rm eq}_{\Lambda} &=& \frac{3\gamma_{_R}-4\gamma_{_M}}{(4-\gamma_{_R})(3-\gamma_{_M})}\rho^{\rm eq}_{_M}, \quad \rho^{\rm eq}_{_M}\approx 
\rho^0_{_M}\left(\frac{a_0}{a_{\rm eq}}\right)^3, 
\label{ceq}
\end{eqnarray} 
where $\rho^0_{_M}$ and $\rho^0_{_\Lambda}$ are the values at the present time $a_0=(1+z)\sim 10^4a_{\rm eq}$. The solution (\ref{rholms}) shows that the term $(\gamma_{_M}/3) \rho_{_M}$ decreases as $a^{-3}$, $\rho_{_\Lambda}$ fails to track down 
$\rho_{_M}$, and becomes positive value approaching to the constant $\rho^0_{_\Lambda}\approx  \tilde {\mathcal C}^{\rm eq}_{\Lambda}$. These results (\ref{rhoms},\ref{rholms},\ref{ceq}) depend on the transition period from radiation to matter epoch. As for the case (ii) that $\rho_{_\Lambda}-\rho_{_R}$ tracking dynamics ends and $\rho_{_\Lambda}$ becomes positive value at sometime $a_{\rm tr}$ around/after the last scattering surface. Discussions and results are similar with substitutions: $a_{\rm eq}\rightarrow a_{\rm tr}$, $\rho^{\rm eq}_{_M}\rightarrow \rho^{\rm tr}_{_M}$ and $\tilde {\mathcal C}^{\rm eq}_{\Lambda}\rightarrow \tilde {\mathcal C}^{\rm tr}_{\Lambda}$ in Eqs.~(\ref{rhoms}-\ref{ceq}).

\comment{The transitions from reheating to radiation $\tilde {\mathcal C}_\Lambda=0$ (\ref{rholrs}) and from radiation to matter ${\mathcal C}^{\rm tr}_\Lambda \not=0$ (\ref{ceq}) occur so that 
the $\rho_{_\Lambda}-\rho_{_R}$ 
tracking dynamics proceeds and fails. More detail numerical analysis are required to study these transitions. 
}

To end this section, we mention $\rho_{_\Lambda}$-dominate epoch in future, when 
$H^2\approx \rho_{_\Lambda}/(3m^2_{\rm pl})$, 
and $\rho_{_M}^H\approx \chi\bar m_{_\Lambda}^2\rho_{_\Lambda}$, 
we recast Eqs.~(\ref{rateeqd}) and (\ref{rhol}) as  
\begin{eqnarray}
\frac{d\rho_{_M}}{dx} + 3 \rho_{_M} &=&  + \gamma_{_\Lambda} 
(\chi\bar m_{_\Lambda}^2\rho_{_\Lambda}-\rho_{_M}),
\label{rhoml}\\
\frac{d\rho_{_\Lambda}}{dx} &=& - \gamma_{_\Lambda}
(\chi\bar m_{_\Lambda}^2\rho_{_\Lambda}-\rho_{_M}),
\label{rholl}
\end{eqnarray} 
where positive $\gamma_{_\Lambda}\equiv\langle \Gamma_M/H\rangle|_{\Lambda}$, 
$\bar m_{_\Lambda}\equiv  (2/3) m_{_\Lambda}/m_{\rm pl}$ and $\chi \bar m_{_\Lambda}^2< 1$. Here we introduce the average mass parameter and rate for this epoch. For $\chi\bar m_{_\Lambda}^2\rho_{_\Lambda}>\rho_{_M}$, namely $\rho^H_{_M}>\rho_{_M}$ in the cosmic rate equation (\ref{rateeqd}) or (\ref{rhol}), asymptotic solutions are slowly time varying
\begin{eqnarray}
\rho_{_\Lambda}\approx \rho^0_{_\Lambda}\left(\frac{a_0}{a}\right)^{
\chi\bar m_{_\Lambda}^2\gamma_{_\Lambda}};\quad \rho_{_M}\approx \chi\bar m_{_\Lambda}^2\gamma_{_\Lambda}\rho_{_\Lambda}.
\label{rholls}
\end{eqnarray}  
It shows that dark energy decreases in time and converts to matter, and the latter tracks down the former. 

\begin{figure}   
\includegraphics[height=5.5cm,width=7.4cm]{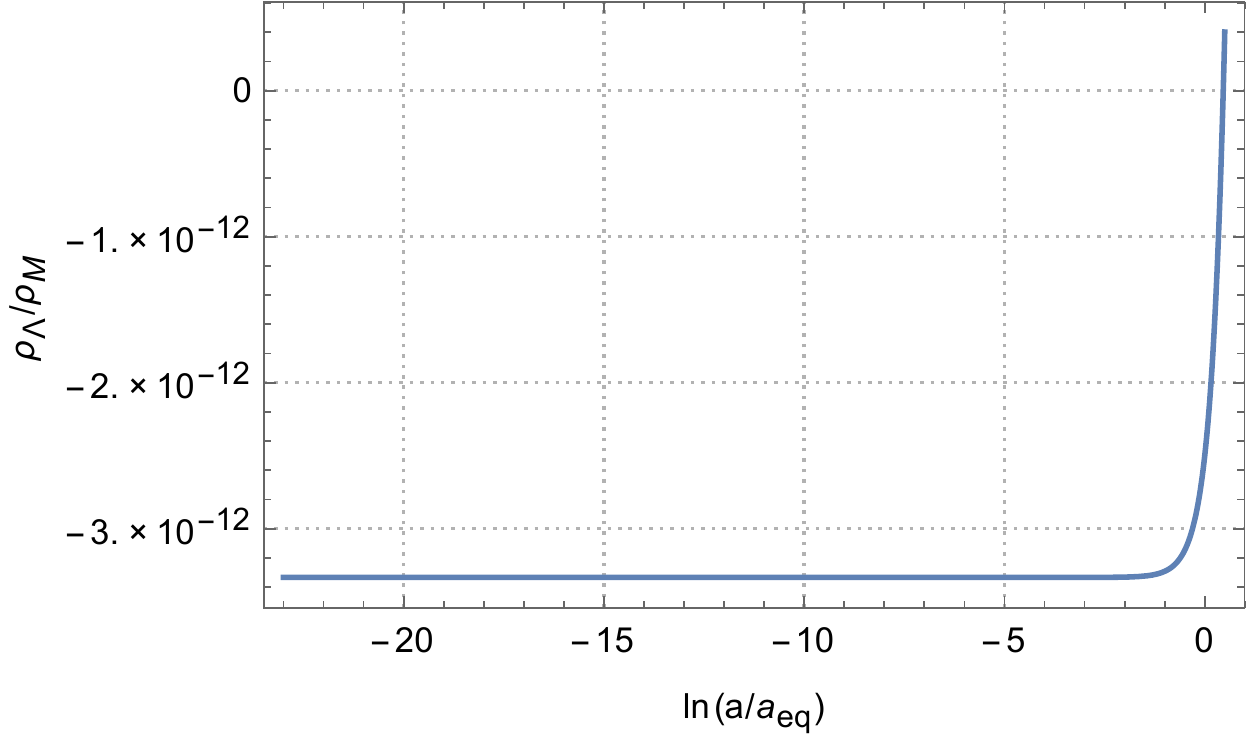}
\includegraphics[height=5.5cm,width=7.4cm]{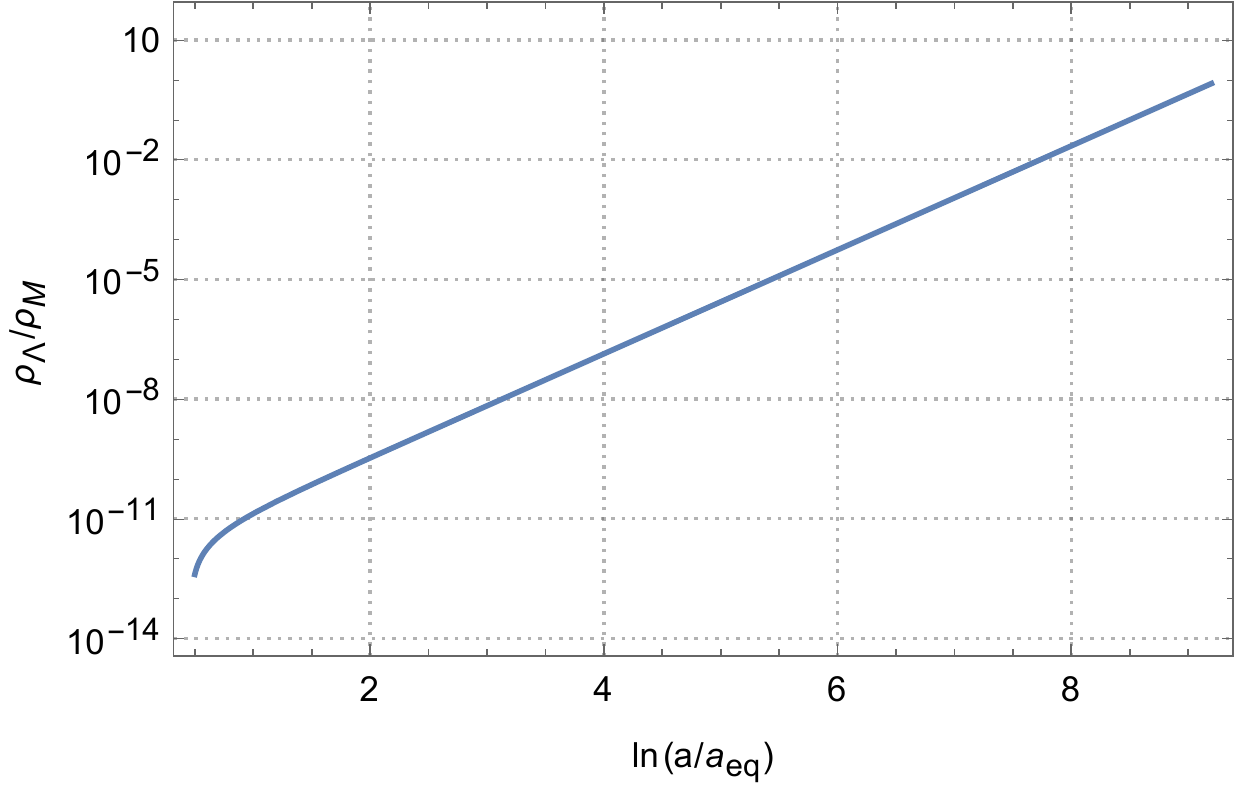}

\includegraphics[height=5.5cm,width=7.4cm]{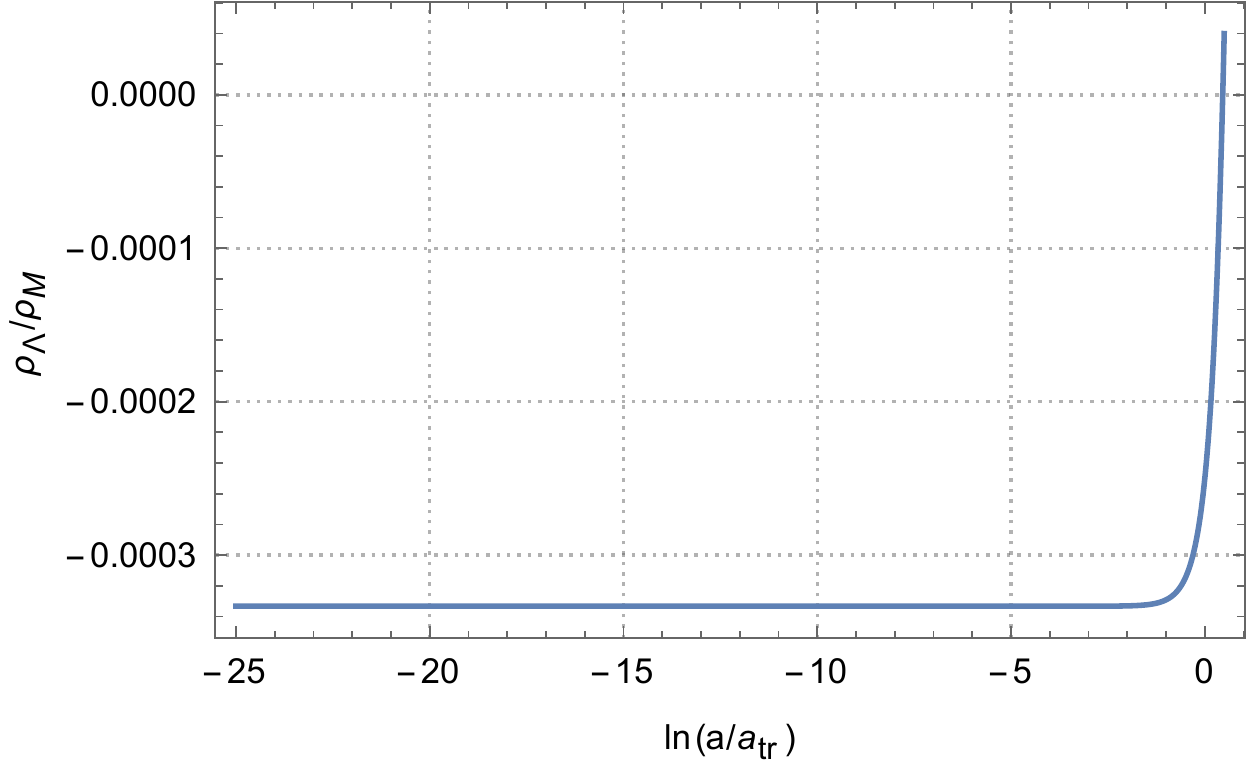}
\includegraphics[height=5.5cm,width=7.4cm]{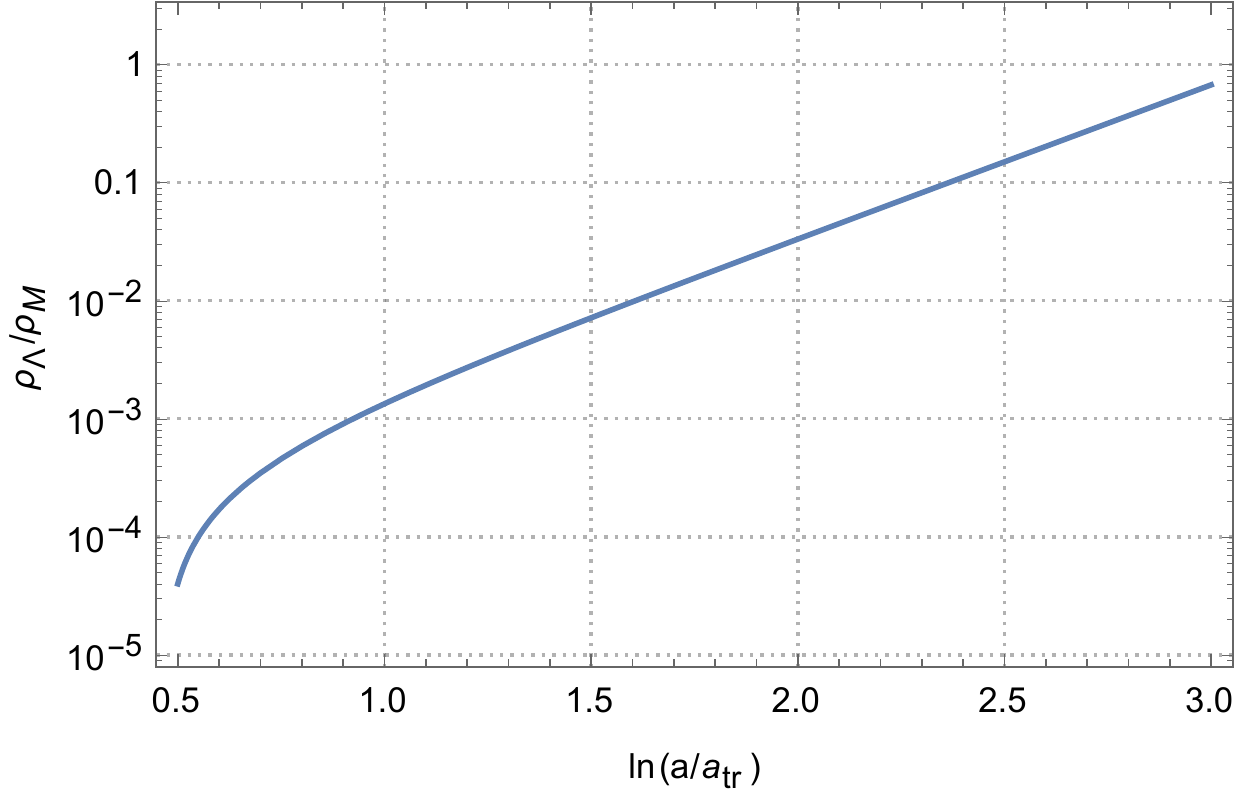}
\caption{The ratio $\rho_{_\Lambda}/\rho_{_M}$ (\ref{gg5+}) is plotted as a function of 
$\ln (a/a_{\rm eq})$, where the scaling factor $a$ 
runs from the reheating end $a_{_R}$, through transition period  $a_{\rm tr}$ to the present time $a_0$, $a_{_R} <a_{\rm tr}<a_0$. It shows that (a) the tracking-down 
behavior: the ratio $\rho_{_\Lambda}/\rho_{_R}$ is a small negative constant $\gamma_{_R}/4$ 
for $\ln (a/a_{\rm tr})<0$; (b) tracking-down failure occurs and dark-energy density $\rho_{_\Lambda}$ becomes positive around $\ln (a/a_{\rm tr})=0$; (c) $\rho_{_M}\sim (a/a_{\rm eq})^{-3}$ (\ref{rhoms}) and $\rho_{_\Lambda}\approx \rho^0_{_\Lambda}$ (\ref{rholms}) constant, 
the ratio $\rho_{_\Lambda}/\rho_{_M}$ increases to 
${\mathcal O}(1)$. The upper panel: $|\gamma_{_M}|\approx |\gamma_{_R}|\sim 10^{-11}$ for the 
case (i) $a_{\rm tr}\sim a_{\rm eq}\sim 10^4a_0$ and the present time $\ln (a_0/a_{\rm eq})\approx 9.2$. The lower panel: $|\gamma_{_M}|\approx |\gamma_{_R}|\sim 10^{-3}$ for the 
case (ii) $a_{\rm tr}\sim 30 a_0$ and the present time $\ln (a_0/a_{\rm tr})\approx 3.4$.}
\label{ccplot}
\end{figure}

\section{Cosmic coincidence of present dark and matter energies}

To discuss the cosmic coincidence, we use the ratio 
$\rho_{_\Lambda}/\rho_{_M}$ which is independent of
the characteristic scales in different epochs. 
We separately discuss two extremal cases:
(i) $a_{\rm tr}\sim a_{\rm eq}\sim 10^4a_0$ or (ii) $a_{\rm tr}\sim 10^2a_0$, when 
the $\rho_{_\Lambda}-\rho_{_M}$ tracking ends and $\rho_{_\Lambda}$ becomes positive. 
In radiation epoch, solution (\ref{rholrs}) shows the ratio $\rho_{_\Lambda}/\rho_{_R} \approx\gamma_{_R}/4$ keeps constant, as $\rho_{_\Lambda}$ tracks down $\rho_{_R}$ from the reheating end $a_{_R}$ to (i)
the radiation-matter equilibrium $a_{\rm eq}\sim 10^{23}a_{_R}$ or (ii) sometime after the last scattering surface $a_{\rm tr}\sim 10^{25}a_{_R}$. 
This tracking dynamics avoids the fine tuning cosmic  
$\rho_{_\Lambda}$ and $\rho_{_R}$ coincidence of the order of 
(i) $(a_{\rm eq}/a_{_R})^4\sim 10^{92}$ or (ii) $(a_{\rm tr}/a_{_R})^4\sim 10^{100}$. Whereas, from the transition time 
(i) $a_{\rm tr}\sim a_{\rm eq}=(1+z_{\rm eq})^{-1}a_0\sim 10^{-4}a_0$ 
or (ii) $a_{\rm tr}\sim (1+z_{\rm tr})^{-1}a_0\sim 10^{-2}a_0$ to the present time $a_0$, solutions (\ref{rhoms}) and (\ref{rholms}) give   
\begin{eqnarray}
\frac{\rho_{_\Lambda}}{\rho_{_M}}
\approx \frac{\gamma_{_M}}{3-\gamma_{_M}} + \frac{(3\gamma_{_R}-4\gamma_{_M})}{(4-\gamma_{_R})(3-\gamma_{_M})}\,\left(\frac{a}{a_{\rm tr}}\right)^{3 -\gamma_{_M}}.
\label{gg5+}
\end{eqnarray}
This ratio consistently approaches the constant $-\gamma_{_R}/4$ when scale factor $a$ traces back to the reheating end $a_{_R}$.
Using (i) $\gamma_{_M}\approx \gamma_{_R}\sim 10^{-11}$ for the case
$a_{\rm tr}\sim a_{\rm eq}\sim 10^4a_0$; (ii) 
$\gamma_{_M}\approx \gamma_{_R}\sim 10^{-3}$ for the case $a_{\rm tr}\sim 10^2a_0$, we plot in Fig.~\ref{ccplot} 
the ratio $\rho_{_\Lambda}/\rho_{_{R,M}}$ varying from $-\gamma_{_R}/4$ 
to ${\mathcal O}(1)$ as a function of the scale factor $\ln (a/a_{\rm tr})$ from the reheating to present time. It shows that 
the cosmic coincidence of the present $\rho^0_{_\Lambda}$ and 
$\rho^0_{_M}$ values appear naturally without any extremely fine-tuning their values at the transition time. 
Namely, in Eq.~\ref(\ref{gg5+}) the ratio $\rho_{_\Lambda}/\rho_{_M}\sim (a/a_{\rm tr})^3$ variation is about (i) ${\mathcal O}(10^{-12})$ or (ii) ${\mathcal O}(10^{-6})$, see the right column of Fig.~\ref{ccplot}.  
The reason is that the matter-dominated epoch of (i) $z_{\rm eq}\sim  10^4$ or (ii) $z_{\rm tr}\sim  10^2$ is much shorter than the radiation dominated epoch of (i) $(a_{\rm eq}/a_{_R})\sim 10^{23}$ or 
(ii) $(a_{\rm tr}/a_{_R})\sim 10^{25}$, 
when the $\rho_{_\Lambda}$ tracks down $\rho_{_R}$ and the ratio $\rho_{_\Lambda}/\rho_{_R}$ is a constant, 
see the left column of Fig.~\ref{ccplot}.  
Otherwise, to reach present $\rho_{_\Lambda}$ and $\rho_{_M}$  observational values of the same order of magnitude, we would have the cosmic coincidence problem of incredibly fine-tuning their 
reheating values $\rho^{\rm RH}_{_\Lambda}$ 
and $\rho^{\rm RH}_{_M}$ at order of (i)
$(10^{-23})^4\times (10^{-4})^3\sim 10^{-104}$ or (ii) $(10^{-25})^4\times (10^{-2})^3\sim 10^{-106}$.

\comment{
We discuss the main points in the present 
scenario for understanding why the cosmological term is ``constant'' in 
the current epoch, and how the fine-tuning problem of cosmic coincidence 
can be possibly avoided.
Equations (\ref{sfriedman}) and (\ref{rateeqd}) give a 
$\rho_{_\Lambda}$ and $\rho_{_M}$ interaction, whose strength depends on evolution epochs and transitions from one to another so that their tracking dynamics proceeds and fails. 
Despite the detailed numerical analysis necessarily required 
for computing the values $\tilde {\mathcal C}_\Lambda=0$ (\ref{rholrs}) and ${\mathcal C}^{\rm eq}_\Lambda \not=0$ (\ref{ceq}), we expect
that the ``continuous'' transition between the reheating epoch 
and the radiation dominated epochs must be different from the 
``discontinuous'' transition between the radiation dominated epoch 
and matter-dominated epoch.
}

\section{Discussions}

Massive pair productions and oscillations on the cosmic horizon lead to 
a massive pair plasma (\ref{apdenm},\ref{prate}). It back reacts on  Friedman equation (\ref{sfriedman}) with matter $\rho_{_M}$ and dark energy $\rho_{_\Lambda}$, via cosmic rate equation (\ref{rateeqd}). 
As a consequence, 
matter and dark energy interact with each other in Universe evolution. The induced dark-energy and matter $\rho_{_\Lambda}-\rho_{_M}$ interacting strength $\langle\Gamma_M/H\rangle$ depends on evolution epochs. 
We study asymptotic solutions for radiation and matter epochs, starting from the reheating end. Because of different epoch transitions, $\rho_{_\Lambda}-\rho_{_M}$ tracking dynamics proceeds in the radiation epoch and ends in matter one. As a result, a slowly varying dark-energy density is of the same order of matter-energy density today. We can avoid the extremal fine-tuning problem of cosmic coincidence. 

Due to the lack of enough knowledge, we have not been able to determine the details of epoch transitions. However, asymptotic solutions (\ref{rhors}), (\ref{rhoms}) and (\ref{rholls}) show 
modified scaling laws in contrast with the counterparts of $\Lambda$CDM. Therefore we consider the following phenomenological model of dark energy and matter interaction.
The Hubble function $E(z)^2=H^2/H^2_0$ can be parametrized 
\begin{eqnarray}
E(z)^2=\Omega_{_R}(1+z)^{4-\delta^G_{_R}} + \Omega_{_M}(1+z)^{3-\delta^G_{_M}} 
+ \Omega_{_\Lambda} (1+z)^{\delta_{_\Lambda}}.
\label{gg6--}
\end{eqnarray}
Here energy densities $\rho_{_{R,M,\Lambda}}$ are in units of the critical density $\rho^0_c=3m_{\rm pl}^2H_0^2$ today, and $\Omega_{R,M,\Lambda}$ are the present values and $\Omega_{_R}+\Omega_{_M}+\Omega_{_\Lambda}=1$. 
Inserting $E(z)$ (\ref{gg6--}) into the dark-energy and matter interacting equation (\ref{sfriedman}), the dark energy term can be obtained as, 
\begin{eqnarray}
\Omega_{_\Lambda}(1+z)^{\delta_\Lambda}= \Omega_{_\Lambda}+ \delta^{M}_{_G}\frac{\Omega_M}{3}\left[(1+z)^{3-\delta^M_{_G}}-1\right]+ \delta^{R}_{_G}\frac{\Omega_R}{4}\left[(1+z)^{4-\delta^R_{_G}}-1\right].
\label{z=1000+}
\end{eqnarray}
Equations (\ref{gg6--}) and (\ref{z=1000+}) give a class of effective interacting dark energy models with two parameters $\delta^M_{\rm G}$ and $\delta^R_{\rm G}$. These modified scaling laws (\ref{gg6--}) were 
also proposed from the view point that time-varying cosmological term $\tilde\Lambda(t)$ and gravitational coupling $\tilde G(t)$ obey scaling laws approaching to their present values ($G,\Lambda$), where Ricci scalar term $R$ and cosmological term $\Lambda$ of classical Einstein gravity are realized \cite{Xue2015} in the spirit of Weinberg asymptotic safety \cite{Weinberg2009} for the quantum field theory of gravity. Based on observational data, the model is 
examined and parameters are constrained in Refs.~\cite{Begue2019} and \cite{Gao2021}, showing it greatly relieves the $H_0$ tensions of the standard cosmology model $\Lambda$CDM.  

We end this article with some remarks. In radiation dominate
epoch, negative dark-energy density $\rho_{_\Lambda}\approx 
\gamma_{_R} m^2_{\rm pl}H^2$ (\ref{rholrs}) follows the ``area law'' $\propto H^2$. In matter dominate epoch, it changes sign at $\rho_{_\Lambda}=0$ in
Eq.~(\ref{rholms}), and approaches a positive constant 
$\rho^0_{\Lambda}\approx 
\tilde {\mathcal C}^{\rm tr}_\Lambda$ (\ref{ceq}). 
The dark energy undergoes these transitions and becomes dominant, converting to matter, and 
matter density $\rho_{_M}$, in turn, tracks down dark energy density
(\ref{rholls}). 
We speculate that such $\rho_{_\Lambda}$-transitions 
should induce the peculiar fluctuations of the gravitational field that possibly imprint on the CMB and matter spectrum, analogously to the integrated Sachs-Wolfe effect. 
 

\section{\bf Supplemental Material: quantum pair oscillation details}\label{dis}
In microscopic time, we plot the Bogoliubov coefficient $|\beta|^2$, the quantum pair density $\rho^{\rm fast}_{_\Lambda}$ and pressure $p^{\rm fast}_{_\Lambda}$, as well as the fast components of Hubble function $H_{\rm fast}$, 
and cosmological term
$\rho^{\rm fast}_{_\Lambda}$. 

\begin{figure*}[h]
\vspace{+3em}
\includegraphics[height=5.5cm,width=7.8cm]{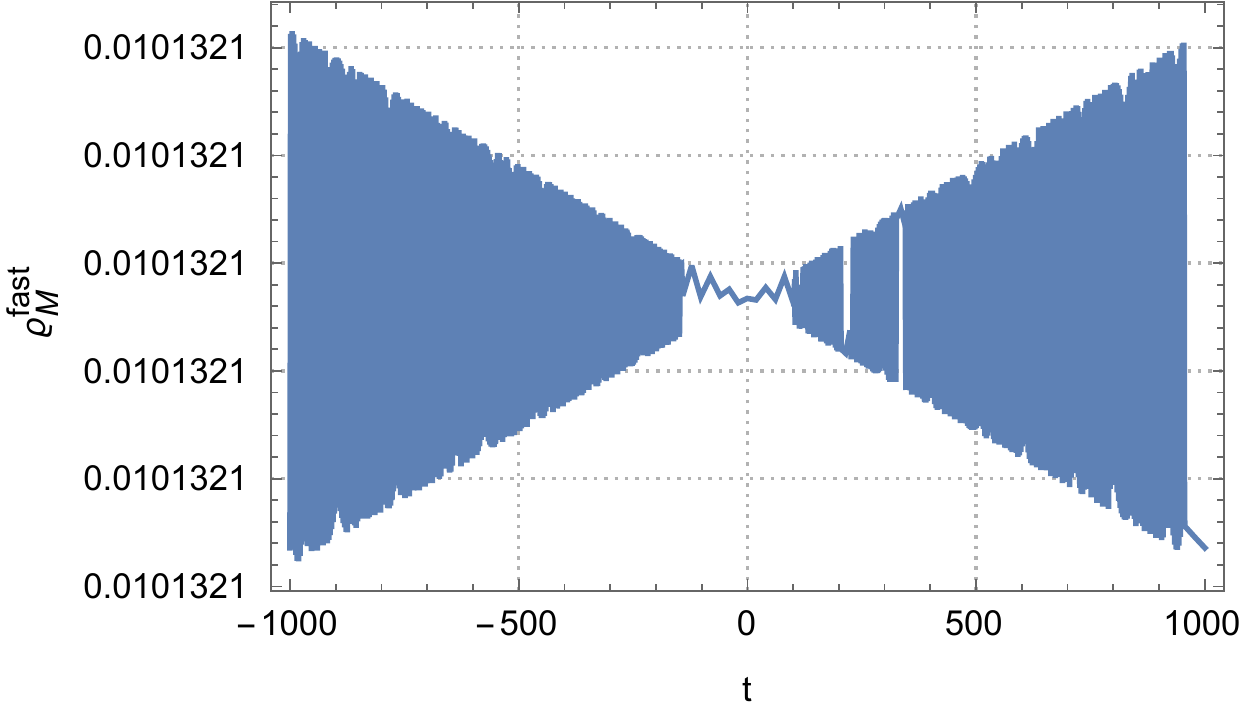}\hspace{0.333cm}
\includegraphics[height=5.5cm,width=7.8cm]{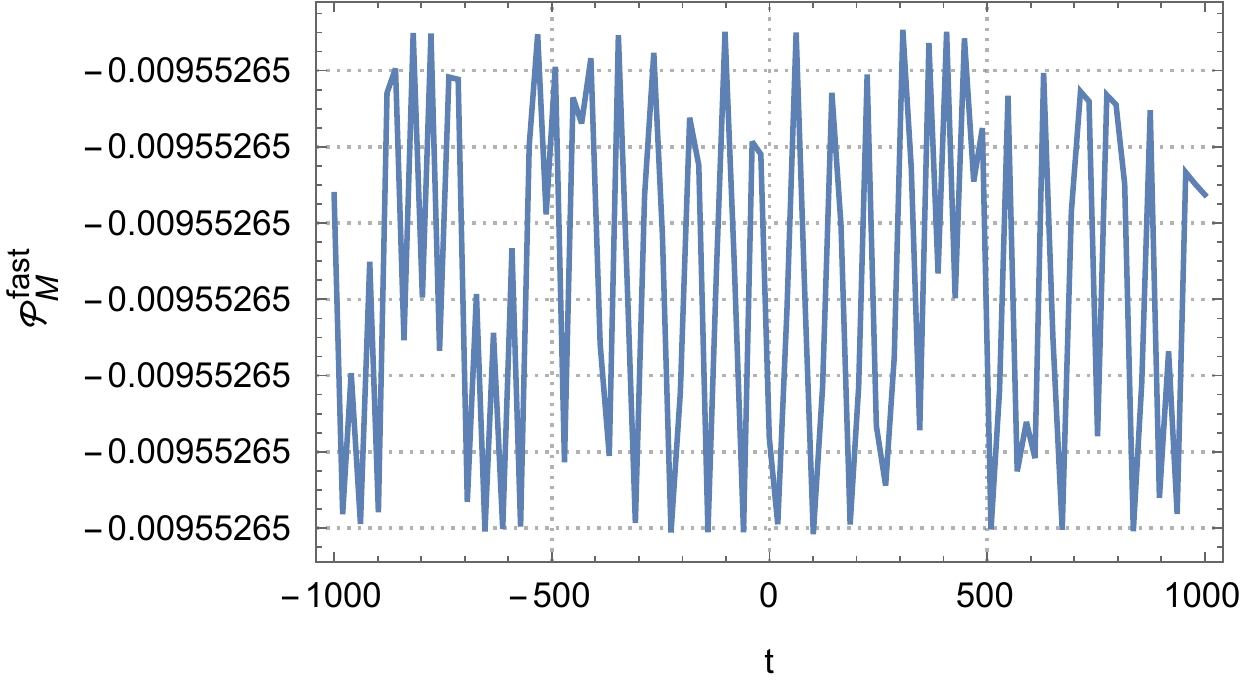}\hspace{0.333cm}
\includegraphics[height=5.5cm,width=7.8cm]{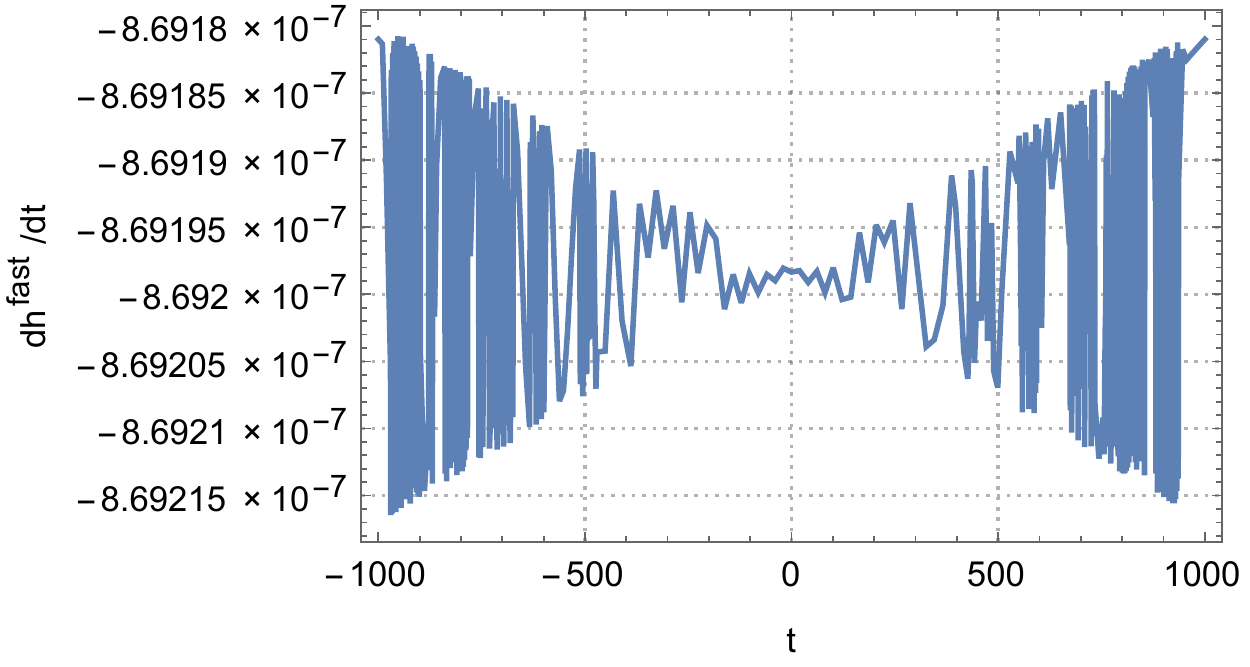}\hspace{0.333cm}
\includegraphics[height=5.5cm,width=7.8cm]{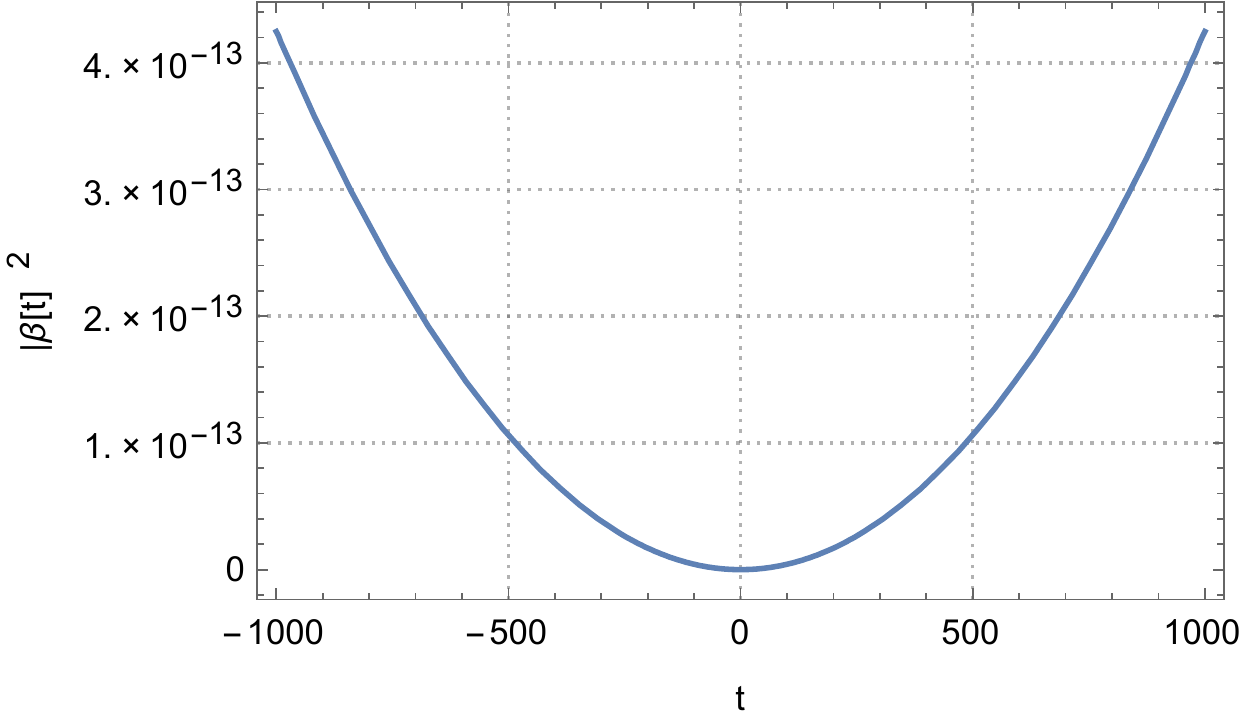}
\includegraphics[height=5.5cm,width=7.8cm]{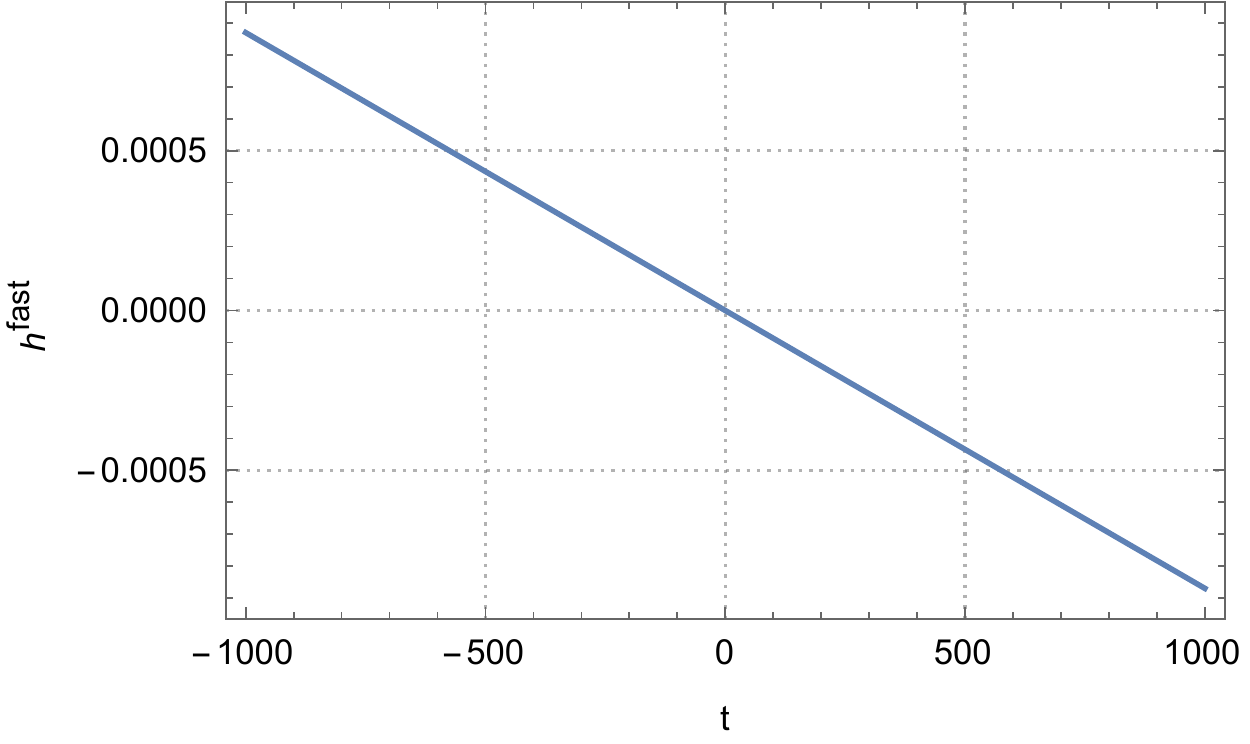}\hspace{0.333cm}
\includegraphics[height=5.5cm,width=7.8cm]{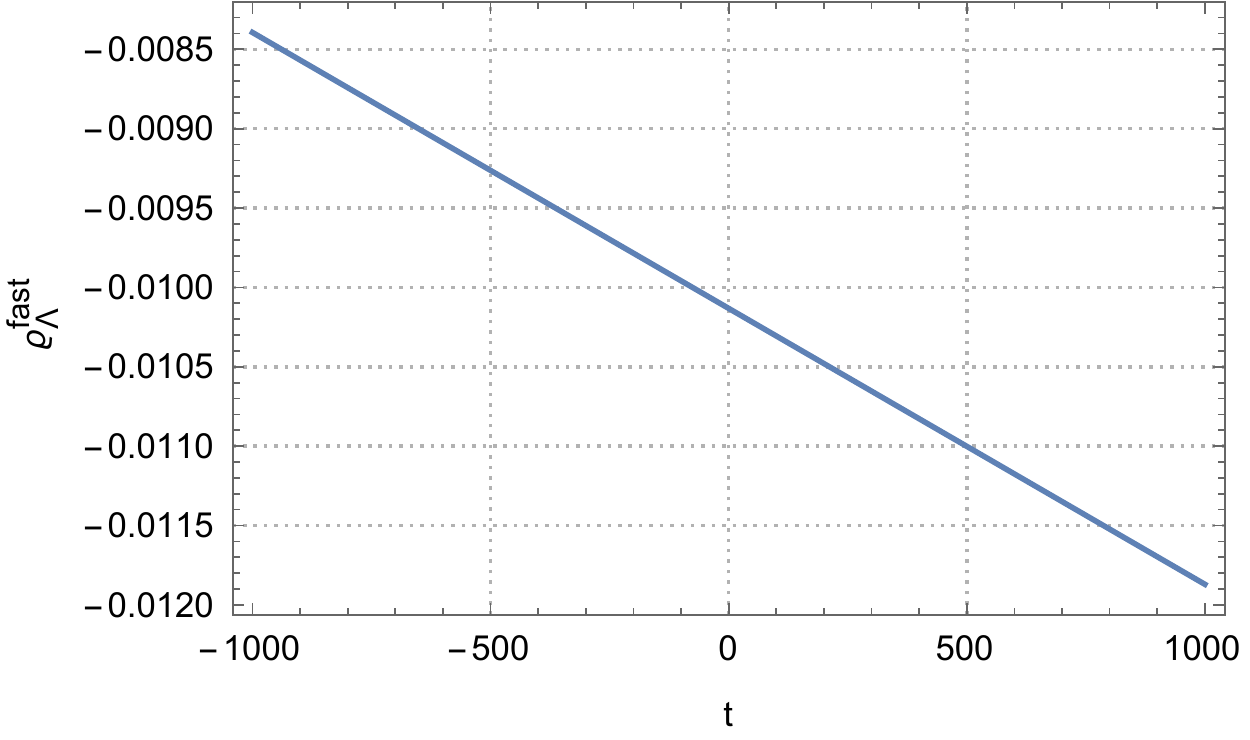}
\caption{Corresponding to Fig.~\ref{osci+}, the details of quantum pair oscillation are shown in microscopic time $t$ in unit of $M^{-1}$. The oscillatory $|\beta(t)|^2$, $h_{\rm fast}$ and $\varrho^{\rm fast}_{_\Lambda}$ structures are too small to see. 
}\label{detailosci1+}
\end{figure*}


\providecommand{\href}[2]{#2}\begingroup\raggedright\endgroup

\end{document}